\documentstyle[11pt,aaspp4]{article}
\def\etal{{\it et al.\/} }

\def\approxlt{\mathrel{\hbox{\rlap{\lower .5ex \hbox {$\sim$}}
	\raise .15 ex \hbox{$<$}}}}
\def\approxgt{\mathrel{\hbox{\rlap{\lower .5ex \hbox {$\sim$}}
	\raise .15 ex \hbox{$>$}}}}

\def\apj{{\it Ap.~J.}}

\def\mnras{{\it M.~N.~R.~A.~S.}}
\newbox\grsign \setbox\grsign=\hbox{$>$} \newdimen\grdimen \grdimen=\ht\grsign
\newbox\simlessbox \newbox\simgreatbox \newbox\simpropbox
\setbox\simgreatbox=\hbox{\raise.5ex\hbox{$>$}\llap
     {\lower.5ex\hbox{$\sim$}}}\ht1=\grdimen\dp1=0pt
\setbox\simlessbox=\hbox{\raise.5ex\hbox{$<$}\llap
     {\lower.5ex\hbox{$\sim$}}}\ht2=\grdimen\dp2=0pt
\setbox\simpropbox=\hbox{\raise.5ex\hbox{$\propto$}\llap
     {\lower.5ex\hbox{$\sim$}}}\ht2=\grdimen\dp2=0pt

\newbox\grsign \setbox\grsign=\hbox{$>$} \newdimen\grdimen \grdimen=\ht\grsign
\newbox\simlessbox \newbox\simgreatbox \newbox\simpropbox
\setbox\simgreatbox=\hbox{\raise.5ex\hbox{$>$}\llap
     {\lower.5ex\hbox{$\sim$}}}\ht1=\grdimen\dp1=0pt
\setbox\simlessbox=\hbox{\raise.5ex\hbox{$<$}\llap
     {\lower.5ex\hbox{$\sim$}}}\ht2=\grdimen\dp2=0pt
\setbox\simpropbox=\hbox{\raise.5ex\hbox{$\propto$}\llap
     {\lower.5ex\hbox{$\sim$}}}\ht2=\grdimen\dp2=0pt

\def\apj{ApJ}

\def\etal{et al}

\begin{document}

\title{Akn 564: an unusual component in the X-ray spectra of NLSy1 galaxies }
  
\author {T.J.Turner \altaffilmark{1,2},
I.M. George \altaffilmark{1, 3},
Hagai Netzer \altaffilmark{4,5}
}

\altaffiltext{1}{Laboratory for High Energy Astrophysics, Code 660,
        NASA/Goddard Space Flight Center,
        Greenbelt, MD 20771}
\altaffiltext{2}{University of Maryland Baltimore County, 1000 Hilltop Circle,
        Baltimore, MD 21250}
\altaffiltext{3}{Universities Space Research Association}
\altaffiltext{4}{School of Physics and Astronomy and the Wise Observatory, 
 Tel Aviv University, Tel Aviv 69978, Israel}
\altaffiltext{5}{The Institute of Physical and Chemical Research (RIKEN), 
Hirosawa, Wako, Saitama 351-0198, Japan}

\begin{abstract}
We present an {\it ASCA} observation of the narrow-line Seyfert 1
(NLSy1)  Arakelian 564.
The X-ray light curve shows rapid variability,  but no evidence 
for energy-dependence to these variations, 
within the 0.6 -- 10 keV  bandpass.
A strong (EW $\sim 70$ eV) 
spectral feature is observed close to 1 keV.  
A similar feature has been observed in 
TON S180, another member of the NLSy1 class of objects, 
 but has not been  observed in 
broad-line Seyfert galaxies. The feature energy 
suggests a large contribution from Fe L-shell lines 
but its intensity is difficult to explain 
in terms of emission and/or absorption from 
photoionized gas. The models which predict most emission 
at 1 keV 
are characterized by extreme values of  column density,  Fe  
abundance and ionization parameter. Models based on gas in 
thermal equilibrium with kT$\sim 1$keV provide an alternative 
parameterization of the soft spectrum. 
The latter may be interpreted as the hot intercloud medium, 
undergoing rapid cooling and producing strong 
Fe L-shell recombination lines. In all cases the physical 
conditions are rather different from those observed 
in broad-line Seyferts. 

The hard X-ray spectrum shows a broad and asymmetric 
Fe K$\alpha$ line of large equivalent width ($\sim 550$ eV)
suggestive of significant emission from the inner accretion disk. 
 The profile 
can be explained by a neutral disk viewed at $\sim 60$ degrees 
to the line-of-sight, contrary to the hypothesis that NLSy1s are 
viewed pole-on. The large EW of this line, the strong 1 keV emission and
the strong optical Fe emission lines all suggest an extreme Fe  
abundance in this and perhaps other NLSy1s.

\end{abstract}

\section{Introduction}

A terminology has developed which names a one extreme 
of the distribution of Seyfert galaxies 
narrow-line Seyfert 1 galaxies (NLSy1s), loosely defined as having 
 FWHM H$\beta <2000$ km/s; \verb+[+O{\sc iii}\verb+]+/H$\beta\ 
< 3$, and  strong Fe{\sc ii} emission. These Seyfert galaxies 
appear to have systematically different, or very extreme X-ray attributes 
compared to the rest of the Seyfert population (which we dub BLSy1s, for 
convenience). Examination of 
X-ray properties across the Seyfert population 
reveals an  anti-correlation between both X-ray index and  
FWHM H$\beta$ (Boller \etal\ 1996; Brandt \etal\ 1997) 
as well as between variability amplitude $\sigma^2_{rms}$ and FWHM H$\beta$ 
(Turner et al. 1999). In addition to the obvious range  of 
index and variability properties across the Seyfert population, it is 
interesting to note those X-ray attributes present in only one subset of
sources.   For example, it is well established that large amounts of 
absorbing material attenuate the X-ray spectra of 
many BLSy1s (with varying degrees of ionization), yet there is little evidence 
of any line-of-sight gas to the nuclei of NLSy1s. 

Theoretical attempts to explain the difference in X-ray properties of NLSy1s
include the idea that these sources accrete  
at a higher rate, given their black hole mass, than the rest of the 
Seyfert population 
(e.g. Pounds, Done \& Osbourne 1995). This may be consistent with
 the observation of Fe K$\alpha$ emission 
from (apparently) highly ionized 
material in NLSy1s (Comastri \etal\ 1998, 
Turner \etal\ 1998) and  the relatively large rapid X-ray variations. 
There were also some attempts (Komossa \& Greiner 1999) to explain the unusually 
steep continuum by dusty warm absorbers.  

\section{ROSAT observations of Akn 564}

Arakelian 564 (IRAS 22403$+$2927, MCG +05--53--012; z=0.024) falls within the 
 NLSy1 ``extreme'' of Seyfert properties (Osterbrock \& Shuder 1982) 
and has relatively strong Ca {\sc ii} and Fe {\sc ii} 
emission lines (van Groningen 1993). 
 Akn 564 shows I(H$\alpha$)/I(H$\beta$)=4.4 (similar to many BLSy1s) and 
 \verb+[+O {\sc iii} \verb+]+/H$\beta \sim 12$. 
Walter \& Fink (1993) suggested that the unusual UV ($\lambda1375 $) 
to 2 keV flux ratio 
in Ark 564 could be due to absorption of the UV continuum but 
X-ray data show no evidence for absorption in excess of the Galactic 
line-of-sight column $N_H(Gal)=6.4 \times 10^{20} {\rm cm^{-2}}$ 
(Stark et al. 1992). 
Unusual  dust-to-gas ratio and different   
 lines-of-sight to the UV  and X-ray continua have also been suggested 
(Walter \& Fink 1993). 
A {\it ROSAT} PSPC pointed spectrum shows  
significant structure indicative of 
strong emission or absorption effects, although the PSPC data  
did not allow a distinction to be made between the presence of 
emission or absorption features (Brandt et al. 1994). 
Interestingly, Crenshaw et al. (1999) detect absorption 
lines in Si{\sc iv}, L$\alpha$, N{\sc v} and 
C{\sc iv}, indicating the presence of at least some material in the 
line-of-sight to the optical emission region. 
Crenshaw et al (1999) find a 
 one-to-one correspondence between Seyfert galaxies that show intrinsic 
UV absorption and X-ray ``warm absorbers,'' indicating that these two 
phenomena are related. Thus the presence of the UV absorber 
suggests we should search for the signatures of X-ray absorption in 
Ark~564. 

\section{ASCA Observation of Akn 564}

{\it ASCA} (Makishima et al. 1996) has two solid-state imaging 
spectrometers (SISs) and two gas imaging spectrometers 
(GISs) at the focus of four co-aligned X-ray telescopes.
This instrument combination yields data over a useful bandpass 
$\sim 0.6 - 10$ keV. Ark 564 was observed over the period 1996 
December 23-24. The data were reduced using standard techniques, as 
described in Nandra et al. (1997a). Data screening yielded effective exposure 
times $\sim 48$ ks in the SIS and $\sim 51$ ks in the GIS instruments.
Ark 564 yielded 1.96$^+_-0.01$ ct/s in SIS0 (0.6-10 keV band).  
In this paper, all energy ranges are given in the observers frame, 
unless otherwise noted. 

The {\it ASCA} data show 
significant flux variability (Fig.~1), with factor-of-two 
changes occurring in the 0.5 -- 10 keV flux, over timescales of 
a few thousand seconds. Construction of a time series with 
256s bins yielded a value for ``excess variance'' 
$\sigma^2_{rms} = 39.3^+_-2.24 \times 10^{-3}$ 
(see Turner et al. 1999, and references therein for a definition 
of this quantity and discussion of results 
for a sample of  NLSy1s). Rapid variations of this amplitude  
are a property of NLSy1s (e.g. Boller et al. 1996). Light curves were also
constructed in two different energy bands, 2-10 keV and 
0.5-2 keV, and the excess variance compared in the two bands. 
This test was prompted by the discovery of energy-dependent variability 
in Ton S180, where the strongest X-ray  variations appear to be observed 
in the soft X-ray band.  
However, we found no evidence for energy-dependent variability in Ark 564.

The X-ray spectrum is remarkable. 
The continuum slope was determined by fitting an absorbed power-law 
to the 0.6-5.0 plus 7.5-10.0 keV band (to avoid confusion due to the Fe K$\alpha$ line). 
Fig~2 shows residuals compared to that  
power-law fit. 
A strong spectral feature is evident at $\sim 1$ keV and 
an excess of emission close to 7 keV, which we know to be 
due to an unmodeled Fe K$\alpha$ line.
The photon index over the (featureless) 1.5-4.5 keV band 
is $\Gamma=2.63^{+0.16}_{-0.03}$ (and probably provides the best 
measure of the underlying continuum slope).

\subsection{The soft X-ray spectrum}

The observation of a strong spectral feature close to 1 keV is 
especially interesting. A similar feature has been  
observed in the bright NLSy1 TON S180 
(Turner, George \& Nandra 1998). 
Features of this nature have not been observed in 
BLSy1s, but something consistent with the same feature has been  
 seen at low S/N in some QSOs (Fiore et al 1998;  
George et al. 1999).  The data (excluding the 5.0 -- 7.5 keV band) 
were successfully modeled 
using a power-law continuum plus Gaussian emission feature.
The best fit gave $\Gamma=2.62^{+0.01}_{-0.02}$ with 
continuum normalization
 $n=1.96^{+0.02}_{-0.04} \times 10^{-2} {\rm photons\ cm^{-2} s^{-1}}$  
at 1 keV; 
no (neutral) absorption was found in excess of the Galactic value, and the 
rest-energy of the Gaussian was $E=0.99^{+0.02}_{-0.04}$ keV with
width $\sigma=0.16^{+0.03}_{-0.02}$ keV and normalization 
$n=1.3^{+0.3}_{-0.1} \times 10^{-3} {\rm photons\ cm^{-2} s^{-1}}$.  
 The equivalent width (EW) of this feature is 
$\sim 70^{+20}_{-10} $ eV  compared to the 
observed continuum level. This fit gave $\chi^2=1332/1261 $   
degrees of freedom ($dof$).
To ensure that the strength of the low energy feature is 
not related to any instrumental effects in the detectors, we  
thoroughly investigated sub-divisions of the data based upon 
instrumental parameters, these efforts are described in the Appendix. 

Intensity-selected spectra revealed no significant variability 
in this spectral feature during the observation. However,  we note 
that the source was in a high flux state for a relatively short time, 
affording no strong constraints on spectral variations.  
 
We attempted some alternative fits where the spectrum of 
Ark~564 is parameterized by a double power-law with 
imprinted absorption edges.  
(We do not discuss models using absorption edges in conjunction with 
 a single power-law, as these were  totally inadequate). 
It is interesting to compare these fits with 
similar ones performed on the PSPC data 
(Brandt et al. 1994) especially in light of an expectation that
we might observe absorption in the X-ray band associated with the  
absorption systems detected in the UV data (Crenshaw et al. 1999).  
We find the {\it ASCA} spectra are not 
adequately fit by a model composed of a double power-law 
with neutral absorber, plus single absorption 
edge. Such a model 
yielded $\chi^2=1546/1260\ dof$ (again excluding the 5.0 -- 7.5 keV band) 
and an edge energy $E=0.60$ keV (i.e. it hit the lower bound allowed 
for the edge energy) with optical depth $\tau=0.73$. 
The fit statistic is $\Delta \chi^2$ of 214 worse than 
that obtained when fitting the feature with a  
Gaussian emission profile (above), 
and the latter does not require the presence of a second power-law 
continuum component. 
 Addition of a second 
edge to the model yields an improvement  
to the fit, giving  $\chi^2=1395/1258\ dof$, with the second edge  energy 
 $E=1.32$ keV and optical depth $\tau=0.15$. 
This fit is $\Delta \chi^2$ of 63 worse than the fit 
utilizing an emission feature and so we do not 
pursue this line of modeling. 
 This conclusion is in agreement with that 
stated in recent work by  Vaughan et al (1999), who find signatures 
of warm absorbers in several NLSy1s, but find the features in Ark~564 
to be better modeled with emission than absorption features. 

The energy of the Gaussian emission component suggests the observed feature 
is a blend of line emission from the K-shells of O and Ne, and the
L-shell of Fe, thus we investigate the agreement between the data and 
realistic physical models which produce strong emission from those  
elements.

\subsubsection{Photoionized Gas}

We attempted to model the emission assuming an  origin in photoionized
gas in equilibrium using models calculated by ION97, 
the 1997 version of the code ION
 (see Netzer 1996). Emission features  could dominate over absorption  
features  if the ionized gas is out of the line-of-sight. 
We allowed the  material to be 
photoionized by the $\Gamma \sim 2.6$ continuum of 
Akn 564.  This steep ionizing continuum yields an emission spectrum 
characteristically  different 
from ``typical'' warm absorber  models, 
like the ones illustrated in several of our
previous publications (Netzer 1996; George et al 1998). In this case it
is characterized by much larger column density and larger $U_X$ than found for 
Seyfert 1 galaxies (cf George et al. 1998).
 We will not expand on this
point which is currently under study. 

The modeling of the blend of lines around 0.8--1.2 keV is 
complicated by the  rich spectrum of
 the Fe {\sc xvii -- xxi}  L-shell lines; 
there are dozens energy levels and hundreds of transitions to be considered.
Great effort has been made by Liedahl and collaborators (e.g.
Kallman et al.  1996 and references therein) to calculate
the atomic data required for such modeling. Previous
versions of ION included only 2-5 Fe recombination lines per ion
(Netzer 1996). We have
extended the list by grouping a detailed line list, generously provided
by D. Liedahl, into 8-12 lines per ion such that the total recombination
emission is identical to the one computed by the more sophisticated
atomic models; this 
allows a reasonable simulation of the Fe recombination spectrum 
 (adequate, given the {\it ASCA} spectral resolution).
These models do not yet contain the contribution due to continuum 
fluorescence (absorption of continuum  radiation by resonance lines) 
which, in some geometries, can approximately double the line intensities. 
The details of this contribution are currently under investigation. 

We used trial models composed of a continuum power-law plus emission 
from ionized gas. Emission models were generated covering
 a column density range of 10$^{22}-10^{23.75}$ cm$^{-2}$; 
ionization parameter $ U_x$ 
in the range  0.1--10.0 (see George et al. 1998 for a definition 
of $U_x$) and covering fraction C$_f=$ 0.1--1.0.
The models that produce the largest summed 0.75--1.2 keV flux 
are those with C$_f \simeq 0.5$, 
column density of about $10^{23.5}$ cm$^{-2}$ and $U_x$ of 5 -- 10.
These models typically give (for solar composition) a 
line blend (summing lines and recombination continua 
in the 0.75--1.2 keV range) with EW  
$\simeq 20$~eV. 
This EW is measured relative to the unobscured continuum and a 
larger EW, by up to a factor 1.2, is measured relative to the observed
continuum (which is somewhat suppressed due to absorption). 
These models did not provide an adequate fit to the data.
 The best fit which was achieved (for solar abundances) 
was that utilizing the pure emission spectrum from the photoionized gas 
which 
yielded $\chi^2 = 1701/1260\ dof$  (excluding the Fe K-shell band) for 
 emission from a column of gas with 
 $N_H=3.0^{+1.6}_{-0.6} \times 10^{23}$ cm$^{-2}$ and $U_x =10^+_-1$.  
Fig.~3 shows the data and model illustrating how little of 
 the observed feature can be explained. In these models production 
of a strong feature at 1 keV leads to 
the expectation of strong features at other energies -- 
ruled out by the {\it ASCA} spectral data. 
We have also explored the possibility that the Fe abundance in this
source is unusually high. This is motivated by the strong
Fe K$\alpha$ line and the strong optical Fe{\sc ii} lines in NLSy1s.
Doubling the Fe abundance results in $\sim 50$\% increase in the EW,
 bringing it close to $\sim 30$~eV, but still falling short of the 
 strength of the observed feature. 

None of the models based upon the emission spectrum from photoionized gas 
provided an adequate fit to the data. 
Although models utilizing simple absorption edges failed to fit the data, 
we considered the effects of absorption by a cloud of 
photoionized gas. A  photoionized cloud produces an absorption 
profile which is obviously different to that of arbitrary absorption  edges 
considered in isolation. We considered both models where  a single 
and a  double power-law continuum were absorbed by 
a cloud of photoionized gas. Fits were performed    
with and without an unconstrained contribution from  the  emission spectrum 
of that gas. However, none of these models produced a satisfactory fit to the 
data.

\subsubsection{Thermal Gas}

Next we considered emission from gas in thermal equilibrium. 
ION does not include a very sophisticated treatment of collisionally excited
transitions in Fe {\sc xvii -- xxiii}; instead it groups those lines into bands
containing very small number of lines per
ion. The overall collisional contribution is maintained assuring energy
conservation. 
This is not adequate  for obtaining the realistic 
emissivity pattern of the Fe L-shell transitions. We have therefore
used the  {\sc mekal} calculations, in {\sc xspec}, to model the 0.8--1.2 keV 
thermal emission, as an alternative to the model from ION. 
We note that  such calculations, while very detailed, do not include
 optical depth
effects that are important for some lines.
Such models provide a reasonable  fit to the bulk of the 
soft feature. A model composed of a continuum power-law plus {\sc mekal}  
component gave a best-fit temperature $kT=1.07^+_-0.04$ keV for the gas 
(assuming cosmic abundances) with $\chi^2=1453/1262\ dof$. While this is 
statistically superior to the fit using a power-law plus ION, 
we note that neither model 
matches the entire flux in the soft hump.  
As expected, for this temperature the dominant Fe 
ions are Fe {\sc xvi -- xx} and the resulting Fe  line complex around
1 keV is very strong. In this model, the flux in the 
thermal component (0.5 -- 2 keV band) is 
$f_{0.5-2}= 2.5 \times 10^{-12} {\rm erg\ cm^{-2}\ s^{-1}}$ 
with a corresponding luminosity 
$L_{0.5-2} = 7 \times 10^{42} {\rm erg\  s^{-1}}$ 
(assuming $H_0=50 {\rm km\ s ^{-1}\ Mpc^{-1}, q_0=0.5} $).  
While the {\it ROSAT} PSPC data were taken at a different epoch, we 
compared those data with this model, to see whether
 there was any inconsistency 
if the model was extrapolated down to 0.1 keV, specifically, to
determine whether the PSPC data fell below the flux of the {\sc mekal}   
component, which would provide significant constraints to our model. 
The PSPC data 
gave  good agreement with the {\it ASCA} model in the 0.1 -0.5 keV band. 
We review the possible origins of  this component in the discussion.

\subsection{Fe K$\alpha$}

A significant Fe K$\alpha$ emission line is evident in the 
{\it ASCA} spectrum (Fig.~2). The 
line profile is asymmetric with a marked red wing, as 
observed commonly in Seyfert 1 galaxies (Nandra et al. 1997b). 
The line was modeled using a broad Gaussian component. This provided 
an adequate parameterization of the line shape, and yielded a 
rest-energy for the line $E = 6.25^+_-0.29$ keV, line width 
$\sigma=1.0^{+0p}_{-0.12}$ keV, normalization $n=9.28^+_-2.11  \times 10^{-5} 
{\rm photons\ cm^{-2}\ s^{-1}}$ and equivalent width $EW=566^+_-128$ eV 
(the $p$ denotes that the parameter hit the hard bound set within the fit). 

The asymmetry of the line lead us to fit a 
disk-line model profile (Fabian et al. 1989). The model
assumes a Schwarzschild geometry and we assumed an emissivity law 
$r^{-q}$ for the illumination pattern of the accretion disk, where 
r is the radial distance from the black hole, and q=2.5 (as found 
for a sample of Seyfert 1 galaxies; Nandra et al. 1997b). We 
assume the line originates between 6 and 1000 gravitational radii 
and we constrained the rest-energy of the line to lie between 6.4 and 7 keV. 
The inclination of the system is defined such that $i=0$ is a disk 
orientated face-on to the observer. 
This model  gave a marginally worse fit than the broad Gaussian model 
for the same number of free parameters 
($\Delta \chi^2=4$ for 1217 degrees of freedom). The rest-energy of the line
was $E = 6.40^{+0.15}_{-0p}$ keV, the inclination was $i=57^{+21}_{-8}$ 
degrees and normalization 
$n=7.09^{+2.02}_{-1.79} \times 10^{-5} {\rm photons\ cm^{-2}\ s^{-1}}$.
The equivalent width was $EW=550^{+128}_{-156}$ eV. 
In both parameterizations of the line, we obtain a flux of 
$\sim 9 \times 10^{-13} {\rm erg\ cm^{-2}\ s^{-1}}$, corresponding to a line 
luminosity $L\sim 2.5 \times 10^{42} {\rm erg\ s^{-1}}$. 

\section{Discussion}

The strong spectral signature in the soft X-ray regime is 
intriguing.  
Starburst activity could explain the presence of a large soft X-ray flux,
and such activity often appears concentrated towards a 
galaxy center. The large X-ray luminosity of this component 
would suggest that it must produce detectable infrared emission if these
X-rays originate from a starburst region. 
In the 0.5 -- 4.5 keV band the luminosity  in the 
soft component alone  is $8 \times 10^{42} {\rm erg\ s^{-1}}$. 
Use of this bandpass allows comparison with a study 
of normal and starburst galaxies (David, Jones \& Forman 1992). Using 
the infrared fluxes reported for Ark~564 (Bonatto \& Pastoriza 1997) 
the estimated X-ray emission from a region dominated by star 
 formation would be 
$L_{0.5-4.5} \sim 6 \times 10^{40} \rm erg\ s^{-1}$, almost two orders 
of magnitude lower than that observed.  
Using again the infrared fluxes from Bonatto \& Pastoriza (1997) 
we calculate indices $\alpha_{60,25}=-0.642$ and 
$\alpha_{100,60}=-0.204$
\footnote{$\alpha_{x,y}=
-log[\frac{F_{\nu}(y)}{F_{\nu}(x)}]/ log[\frac{\nu(y)}{\nu(x)}]$  }.
We compared these indices with those 
calculated for samples of Seyfert galaxies and starburst dominated galaxies 
(Miley, Neugebauer \& Soifer 1985). 
The 60$\mu$m curvature is not consistent with that observed in 
starburst galaxies (c.f. Miley, Neugebauer \& Soifer 1985; their Fig.~4). 
Thus the infrared luminosity and spectrum indicate that 
starburst activity does not make a significant contribution to 
the observed X-ray luminosity in Ark~564. 

The optical image of Ark~564 does show extent with ridges of enhanced emission
with a red optical color (Arakelian, 1975). The extended 
optical features give Ark~564 a diameter of $\sim 30''$, with $12''$ 
separating the nucleus from one bright optical region to the SE. 
The {\it ROSAT} HRI image of this source 
shows no significant extended emission coincident with the optical 
enhancements. The 
{\it ROSAT}  HRI data  provide an upper limit such that 
the emitting gas must exist within the inner $\sim 3$ kpc 
of the nucleus,  
much stronger constraints will be obtained by AXAF images 
in the near future. 
(Unfortunately faint extent would  
be very difficult to confirm with the HRI, because the instrument suffers 
some scattering of point-source photons into a halo outside of the 
point-spread-function.)  
An extended thermal source would not show rapid time variability.
While we see strong X-ray flux variability in 
Akn 564 it is 
currently unclear whether the flux in the spectral feature 
itself varies. 
The current constraint, that the hot gas lies within a 3 kpc radius, 
leaves many possibilities open to discussion. However, 
no such component is observed in normal galaxies or 
BLSy1s, thus it seems more compelling to discuss possible origins 
associated with the nucleus, rather than 
further out  (such as gas in the galactic halo).

The {\it ASCA} data show the spectral feature at 1 keV is 
not produced by emission or absorption from 
gas in photoionization equilibrium. 
An alternative is that the gas is in thermal equilibrium. 
The {\sc mekal}  model can be extrapolated to give a bolometric luminosity 
in that component (over $\sim$ 0.1-10 keV, although contributing little 
above 2 keV) which is $10^{43} {\rm erg\ s^{-1}}$ 
(c.f. the nuclear luminosity $L_{2-10} \sim 5 \times 10^{43} 
{\rm erg\ s^{-1}}$).
The temperature determined from our modeling is $kT=1$ keV.
This can be  explained if the gas exists in a 
location where some  heating mechanism dominates over 
photoionization heating. This could be  part of the nucleus 
obscured from the central radiation field or a location where
the density is high enough that collisional cooling is 
able to maintain a low level of ionization. This would give rise to 
strong line emission. 

We consider an alternative origin as the hot intercloud medium (HIM)
confining the clouds in either the broad line region (BLR) 
or narrow line region (NLR).  
Such gas has been postulated in early studies of AGN. 
Krolik, McKee and Tarter (1981) discuss the expectation that 
emission line clouds must be confined by a hot ($10^7$ -- $10^9$K) 
medium (also see Netzer 1990 and references therein).
This medium has also been discussed in connection with 
warm absorber  gas in AGN, by
Reynolds and Fabian (1995). 
The situation regarding NLSy1 is of particular interest since 
 the HIM is 
likely to be at relatively low temperature 
due to the inefficient Compton heating (because of 
the very soft X-ray continuum)
and the large Compton cooling (due to the strong UV bump). Such material, if
in equilibrium, does not emit strong emission lines since it is fully
ionized. 
However, the HIM may be unstable. NLSy1s 
are known to have large flux variations and it is conceivable that
the current observed state of Akn~564 represents a low-phase of activity. 
 Previous higher luminosity  phases might have 
resulted in full ionization of the HIM which, at the time of 
the {\it ASCA} observations was 
undergoing intense cooling and recombination. The recombination time depends
on the density and temperature. Given a BLR location, and a temperature
of $10^6$K, the recombination time is short enough to 
produce a strong recombination 
spectrum. Full modeling of such a time-dependent component is beyond
the scope of this paper. We note, however, that 
non-equilibrium situations are likely to be important in some cases 
involving rapidly varying sources (Nicastro et al. 1999). 

The assumption that there is a strong thermal component in
Akn~564  can be put to  observational tests. First, this component is unlikely
to respond on short time scales to large continuum variations. Its relative
contribution at high phases of the source should  decrease inversely with the
continuum brightness. Second,  high resolution spectral observations, like
those expected with several of the coming X-ray missions ({\it Chandra}  and
{\it XMM}), will provide clear diagnostic tools to allow the identification of
the thermal plasma and its separation from any photoionized plasma. 
If thermal gas is indeed found in high 
resolution observations of Akn~564, it will provide yet another clue to the
unique spectral appearance of NLSy1 galaxies.

Turning now to the properties of the  Fe K$\alpha$ line in Ark~564, we note an
asymmetric shape  as observed in BLSy1s. 
The pronounced red wing  indicates that the innermost disk is emissive. 
We find an acceptable  explanation of the Fe K$\alpha$ line
 as emission from an accretion  disk composed of
neutral material, but viewed at $\sim 60$ degrees to the line-of-sight.  This
inclination angle contradicts the popular hypothesis that
NLSy1s are Seyfert nuclei viewed pole-on.
Data from Ark~564 and Ton S180 indicate that the Fe K$\alpha$ 
lines may have systematically large  equivalent 
width in NLSy1s, compared to BLSy1s. The unusual spectral 
feature observed at 1 keV in these two NLSy1s 
is consistent with emission lines from the L-shell of Fe. 
Ark~564 and Ton S180 show strong 
 optical Fe{\sc ii} emission, as do NLSy1 galaxies in general.
 These facts suggest that overabundance of 
Fe may be a property of NLSy1s.

\section{Conclusions}

An {\it ASCA} observation of the NLSy1 galaxy Ark 564 reveals 
rapid X-ray variability, although we find no evidence for 
energy-dependent flux changes. Ark~564 has a 
complex X-ray spectrum. A strong feature is observed close to 1 keV, 
which is not easily attributed to emission or absorption by 
photoionized gas or emission from gas in 
thermal equilibrium. 
A similar feature was previously observed in 
Ton S180, these features may be characteristic of NLSy1s and important 
to understanding conditions in these sources. 
The emission may indicate the hot intercloud medium in the BLR or NLR 
is currently in a cooling phase in Ark 564.
The hot intercloud 
medium may have a characteristically different temperature in NLSy1s to 
the rest of the Seyfert population,  affording a visible signature 
of the gas in the soft X-ray spectrum of NLSy1s, but not BLSy1s. 
 Separation 
of Ne and Fe lines over the 0.8--1.2 keV range with {\it XMM} or 
 {\it Chandra} grating observations will enable us to test this explicitly. 

Ark~564 has a broad and asymmetric 
Fe K$\alpha$ line of large equivalent width, $\sim 550$ eV. The line profile 
is consistent with emission from a neutral disk inclined at 60 degrees 
to our line-of-sight which  argues against models inferring NLSy1s 
are viewed pole-on. 

\section*{Appendix}

We investigated whether the strength and shape of the low
energy feature in Ark 564 changed as a function of any instrumental 
parameter. First we examined each of the SIS detectors. 
As the lower level for  
event discrimination  changed during the observation 
we accumulated two spectra for each SIS, one for each 
discriminator level.  A separate response matrix was 
generated for each SIS and for each discriminator level. 
These four SIS spectra were then fit to the power-law model, 
and the data/model ratios are shown in Fig.~4. The SIS0 and SIS1 
data are shown separately. It is clear that the spectral feature 
is evident in both, showing consistent strength in each. 
 Fig.~4 shows a comparison between 
spectra from each discriminator level, for the two SIS instruments, 
demonstrating the close consistency between each pair. As no 
significant difference was found,  henceforth we 
show only the average SIS0 spectrum and average SIS1 spectrum. 
 The known discrepancy between SIS0 and SIS1 data 
is evident. SIS1 data lie 10\% lower than SIS0 data below an 
energy of $\sim 0.8$ keV. However, the peak and turnover energy 
of the low energy feature are clearly seen in SIS0, therefore the 
overall shape and strength of the feature is not an artifact of problems with 
SIS1. We found no difference in our conclusions considering SIS0 
data alone, and as there are small uncertainties in all instruments, we 
consider the best approach to be that which we present, analysis based upon 
simultaneous fits to the four {\it ASCA} instruments. 

There are also documented problems with the SIS response matrix 
generator for 2-pixel events in cases when the event threshold is high.
There has been some speculation that a more reliable 
analysis can be achieved by consideration of the grade 0 events 
alone (for which accurate response matrices can be generated regardless of 
event threshold),  
 therefore we repeated our analysis using only the grade 0 events.  
 Fig.~4 shows the data/model ratio for SIS0 and SIS1 
separately, for spectra accumulated from grade 0 events alone. 
We find no dependence of the strength of the feature on 
grade of event. 

The data were also subdivided based upon the CCD temperature during the 
observation. We found no changes in the strength of the low 
energy feature as the CCD temperatures changed. To illustrate 
this point we show data/model ratios to the power-law model, for
two SIS1 spectra; the first accumulated for a mean instrument 
temperature of -60C, the second for mean temperature -61 C 
(close to the full range of temperature during the observation).

In conclusion, we have screened and subdivided 
the data based upon a large number of 
different criteria, but found no instrumental effects which 
falsely enhanced the strength of the low energy feature. 
Our conclusions are not 
dependent on the small residual 
uncertainties in the {\it ASCA} calibration. 

\section{Acknowledgements}
We are grateful to the {\it ASCA} team for their operation of the satellite. 
 This research has 
made use of the NASA/IPAC Extragalactic database,
which is operated by the Jet Propulsion Laboratory, Caltech, under
contract with NASA; of the Simbad database, 
operated at CDS, Strasbourg, France; and data obtained through the 
HEASARC on-line service, provided by NASA/GSFC. 
This work was supported by NASA Long Term Space Astrophysics grant 
NAG 5-7385.

\section*{References}
{\noindent}Arakelian, M.A. 1975, SoByu 47, 3 \\ 
Boller et al 1996A\&A, 305, 53 \\ 
Bonatto, C.J., Pastoriza, M.G. 1997, \apj\ 486, 132 \\
Brandt, W.N., Fabian, A.C., Nandra, K., Reynolds, C.S., Brinkmann, W. 1994, 
	MNRAS, 271, 958 \\
Brandt et al 1997 MNRAS, 285, 25 \\
Comastri et al 1998, A\&A, 333, 31 \\
Crenshaw, D.M., Kraemer, S.B., Boggess, A., Maran, S.P., 
	Mushotzky, R.F., Wu, C-C.   1999, ApJ 516, 750 \\
David, L.P., Jones, C., Forman, W. 1992, \apj\ 388, 82 \\
Fabian,A.C., Rees, M.J., Stella, L., White, N.E. 1989, MNRAS 238, 729 \\
Fiore,F., et al. 1998, MNRAS 503, 607 \\
George, I.M., Turner,T.J., Netzer, H., Nandra, K., Mushotzky, R.F., 
	Yaqoob, T. 1998, ApJS 114, 73 \\ 
George, I.M., et al 1999, ApJ, in prep \\
Kallman, T. R., Liedahl, D., Osterheld, A., Goldstein, W., Kahn, S. 1996, 
  \apj\ 465, 994 \\
Komossa, S., Greiner, J. 1999, (Astro-ph9810105) \\
Krolik, J.H., McKee,C.F., Tarter,C.B. 1981, \apj\ 249, 422 \\
Makashima, et al. 1996, PASJ 48, 171 \\
Miley,G.K., Neugebauer, G., Soifer, B.T. 1985, \apj\ 293, 148 \\
Nandra, K., George,I.M., Mushotzky, R.F.,Turner,T.J.,Yaqoob, T. 
 	1997a, \apj\ 476, 70 \\
Nandra, K., George,I.M., Mushotzky,R.F., Turner,T.J., Yaqoob,T. 
 	1997b, \apj\ 477, 602 \\
Netzer,H. 1990, in {\it Active Galactic Nuclei}, by Woltjer, Netzer \& 
 Blandford, SAA-FEE series, Courvoisier and Mayor eds (Berlin, Springer) p57 \\
Netzer, H. 1996, \apj\ 473, 781 \\
Netzer, H. 1999, in prep \\
Nicastro, F., Fiore, F., Perola, G.C., Elvis, M., 1999, ApJ 512, 184 \\
Osterbrock, D.E., Shuder, J.M., 1982, ApJS 49, 149 \\
Pounds, K.A., Done, C., Osbourne, J.P., 1995, MNRAS 277, L5 \\
Reynolds, C., Fabian, A.C. 1995, \mnras\ 273, 1167 \\
Stark, A.A., Gammie, C.F., Wilson, R.W., Bally, J., Linke, R.A., Heiles, C., 
	Hurwitz, M., 1992, ApJS 79, 77 \\
Turner, T.J., George, I.M., Nandra, K. 1998, ApJ, 508, 648 \\
Turner, T.J., George, I.M., Nandra, K. 1999, ApJ, sched. Nov 1 issue, 
	astro-ph/9906050 \\
Van Groningen, E. 1993, A\&A 272, 25 \\
Vaughan, S., Reeves, J., Warwick, R., Edelson, R. 1999, MNRAS in press \\
Walter, R., Fink, H.H., 1993, A\&A 274, 105 \\

\newpage
\begin{figure}[h]
\plotfiddle{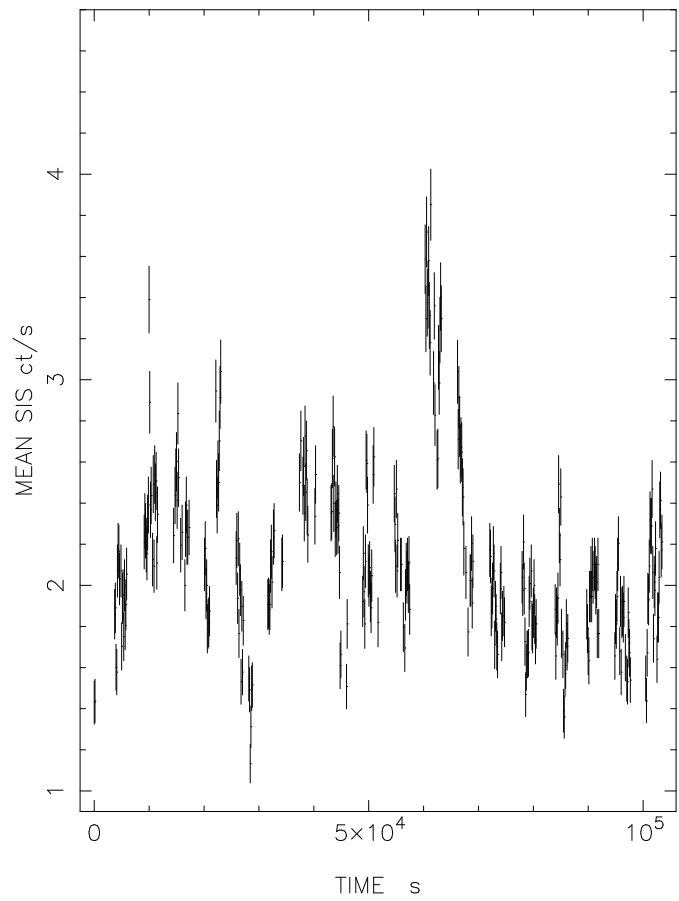}{10cm}{0}{50}{50}{-200}{-20}  
\caption{ {\it The ASCA light curve in the 0.5-10 keV band, 
 based upon the combined SIS data, in 128 s bins. } }
\end{figure}

\begin{figure}[h]
\plotfiddle{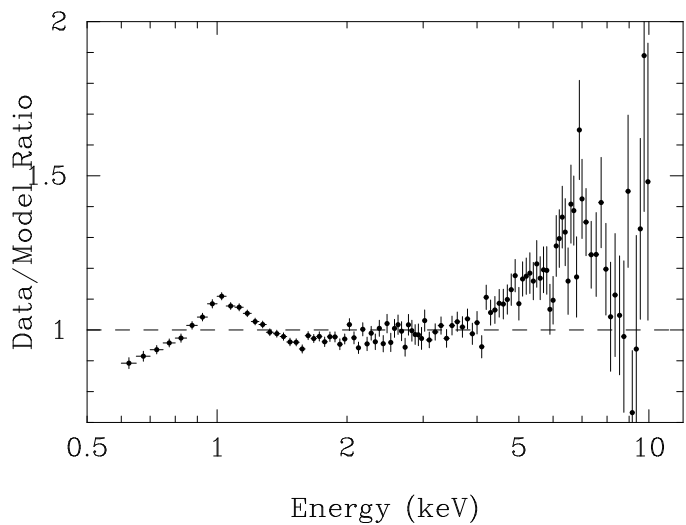}{10cm}{0}{50}{50}{-200}{0}  
\caption{ {\it  The data/model ratio (combined SIS and GIS data) 
compared to a power-law continuum 
fit to the 0.6-5.0 plus 7.5-10.0 keV ASCA data, with the 5.0 -- 7.5 
keV data overlaid for comparison.  
 Residual counts due to emission from  Fe K$\alpha$  are  
evident, as well as the 
strong feature close to  1 keV. } }
\end{figure}

\begin{figure}[h]
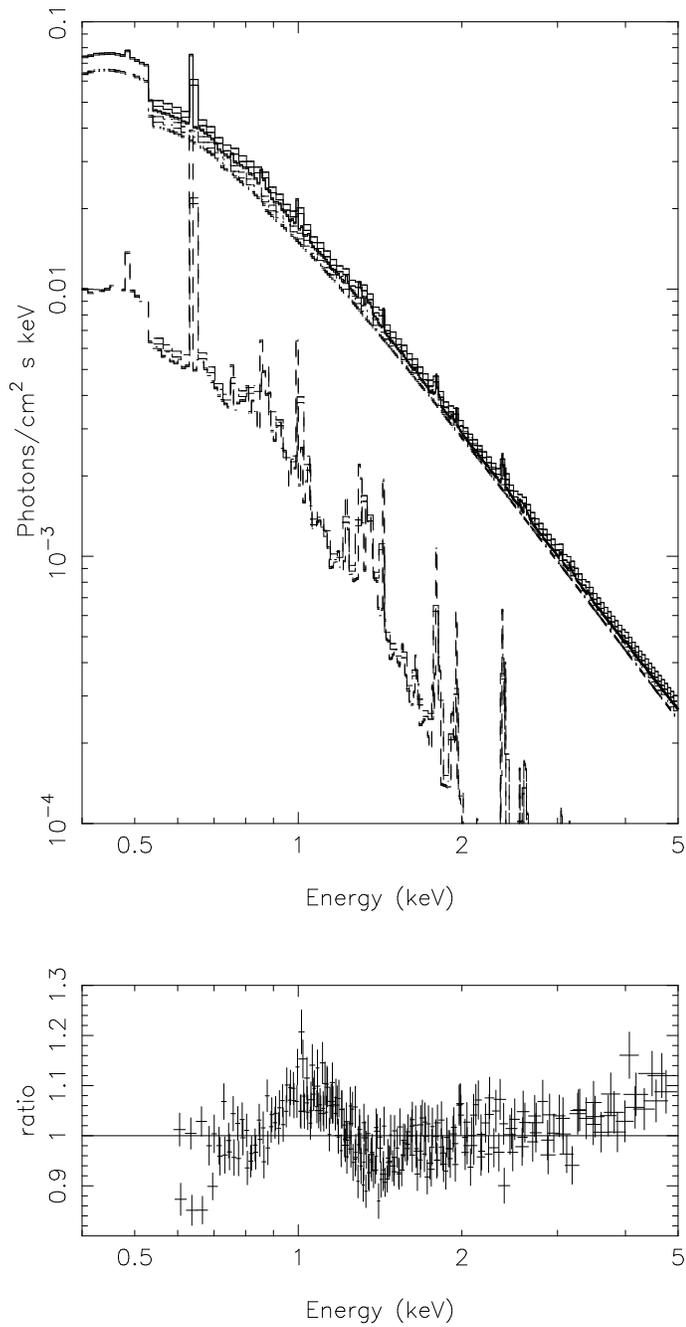

\plotfiddle{closeup_ionmo.vps}{4.5cm}{0}{50}{50}{-200}{0} 
\plotfiddle{closeup_rat.vps}{4.5cm}{0}{50}{50}{-200}{-220}
\caption{ {\it Top: The model for the best fit using the photoionized 
gas, as detailed in \S 3.1.1. Bottom: The data/model ratio with 
each of the four instruments plotted   }} 
\end{figure}

\begin{figure}[h]
\plotfiddle{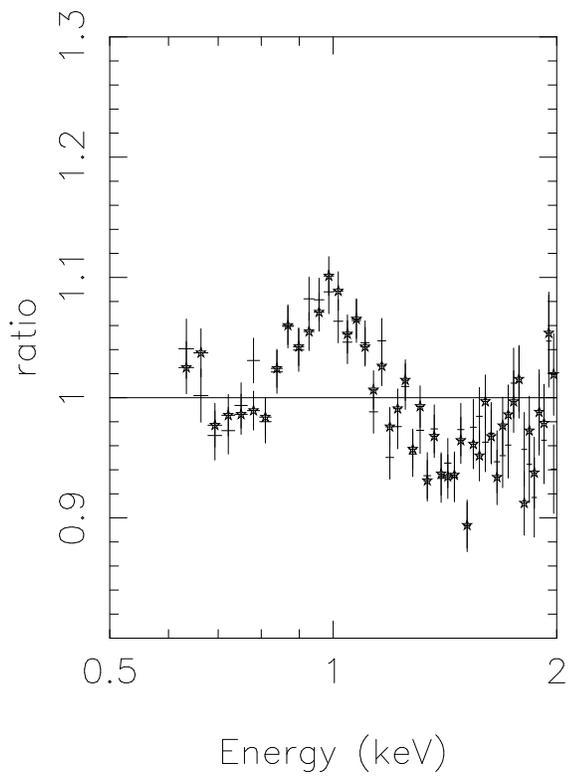}{5cm}{0}{50}{50}{-300}{-20} 
\plotfiddle{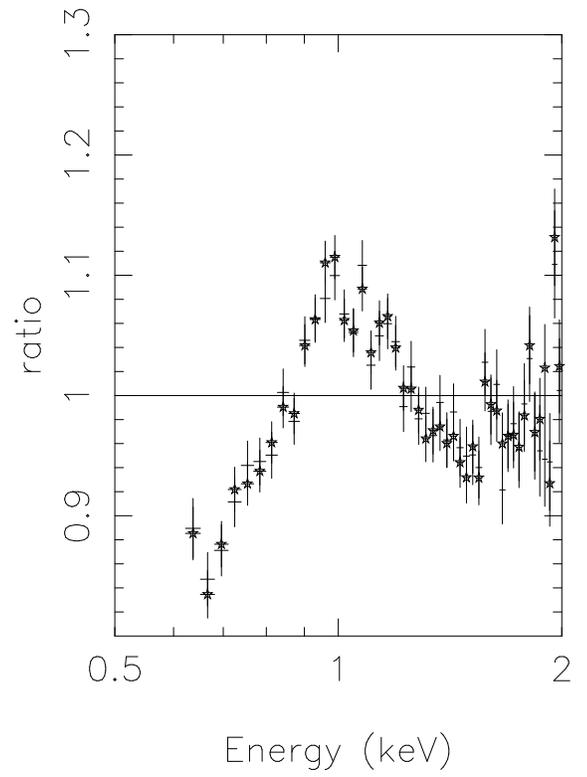}{5cm}{0}{50}{50}{0}{+140}
\caption{ 
{\it The data/model ratio as in Figure 2, except that 
data are shown for each individual SIS and each  
level discriminator used during the observation. 
Left: SIS 0,  stars are data where S0LVDL=135, the crosses  are 
S0LVDL=115. 
Right: SIS 1, the stars show S1LVDL=146, the crosses
 are S1LVDL=125
 } } 
\end{figure}

\begin{figure}[h]
\plotfiddle{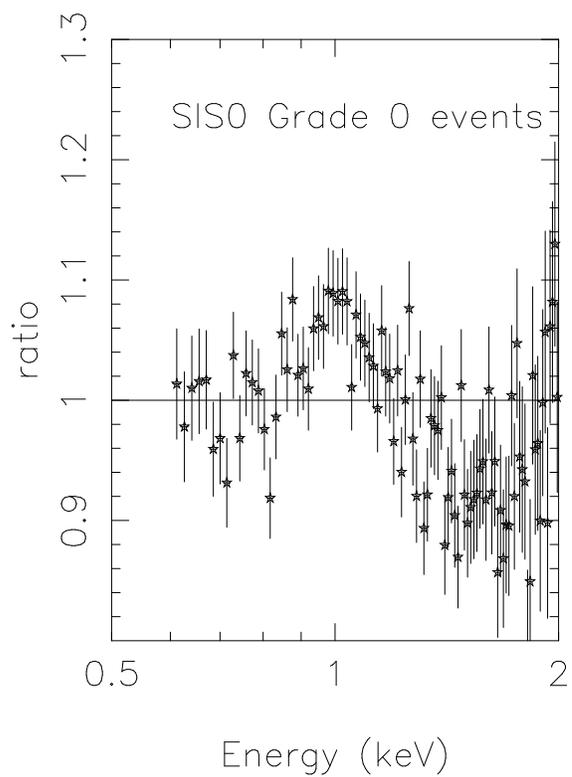}{5cm}{0}{50}{50}{-340}{-150} 
\plotfiddle{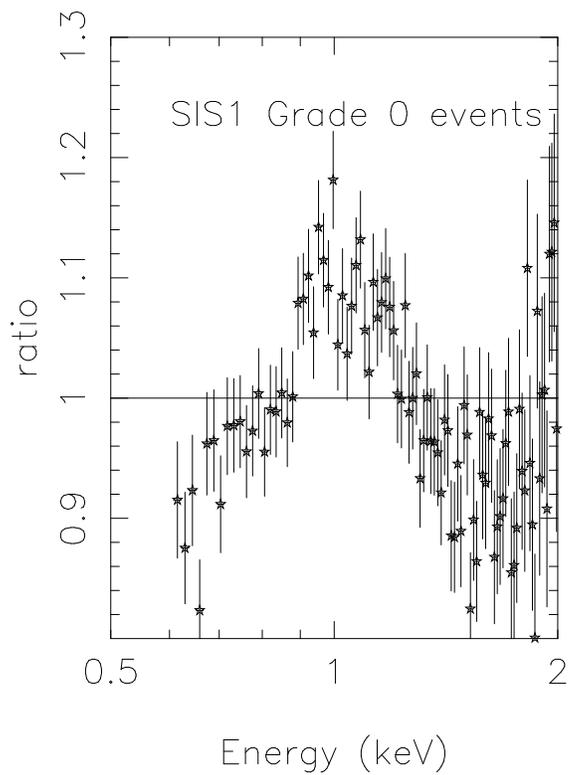}{5cm}{0}{50}{50}{-20}{+10}
\plotfiddle{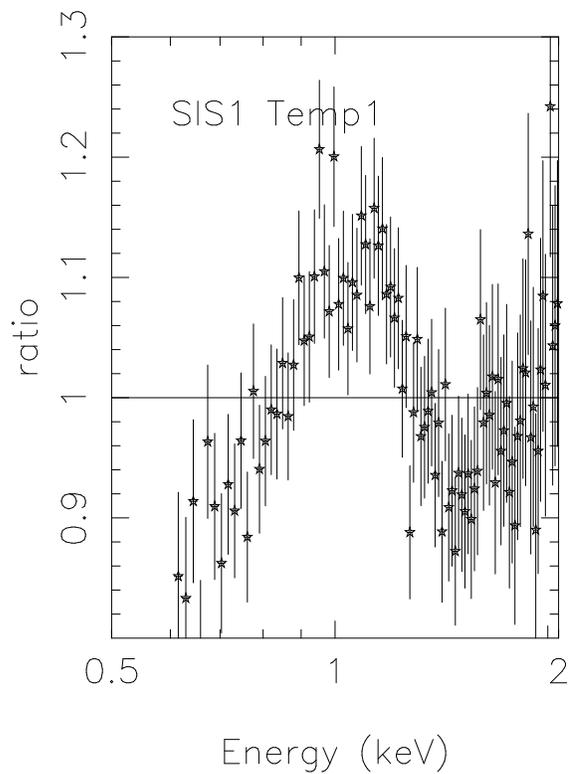}{5cm}{0}{50}{50}{-340}{-200}
\plotfiddle{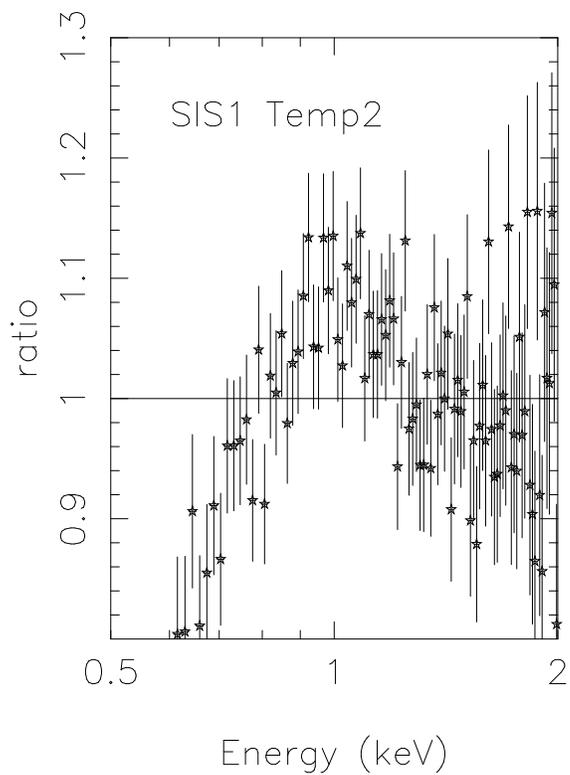}{5cm}{0}{50}{50}{-20}{-40}
\caption{ {\it Data/model ratio as in Figure 2. Top: 
 Grade 0 data from SIS 0 and SIS 1. 
Bottom:  Data/model ratio for SIS 1  subdivided by 
CCD temperature, T1=-60 C (left) and T2=-61 C} (right)}
\end{figure}

\end{document}